\documentclass[11pt, a4paper, onecolumn]{article}
\usepackage[dvips]{graphicx}
\usepackage{graphicx, color}
\usepackage{dcolumn}
\usepackage{bm}
\usepackage{times}
\usepackage[english]{babel}
\usepackage{t1enc}
\usepackage{cite}
\begin{document}

\title{Simulating Bell inequalities violation with classical optics encoded qbits}       
\author{Matías A. Goldin, Diego Francisco, Silvia Ledesma}       
\date{March 2009}          
\maketitle

\begin{abstract}

We present here a classical optics device based on an imaging architecture as analogy of a quantum system where the violation of the Bell inequality can be evidenced. In our case, the two qbits entangled state needed to obtain non classical correlations is encoded using an electromagnetic wave modulated in amplitude and phase. Computational states are represented in a way where each one of the two qbits is associated with two orthogonal directions in the input plane. In addition, unitary operations involved in the measurement of the observables are simulated with the use of a coherent optical processor. The images obtained in the output of the process, contain all the information about the joint, marginal and conditional probabilities. By measuring the intensity distribution in the image plane we evaluate the mean values of the simulated observables. The obtained experimental results show, in an illustrative manner, how some correlations of Clauser-Horne-Shimony-Holt type exceed the upper bound imposed by the local realism hypothesis as a consequence of the joint effect of entanglement and two-particle interference. 

\end{abstract}

\section{Introduction}  

In this paper we present a classical analogy of some well known quantum experiments involving Bell inequalities violation. The analogy between quantum mechanics and classical optics has been recently explored \cite{Cerf, Spr98, Spr01, Bhatt, Puen, FranHad, FranQRW, FranTel}. The main idea is to exploit the wave nature of the electromagnetic field in order to represent the quantum state of one or more particles. In this representation, the probability amplitude of ocurrence of each state of a basis is associated to the complex amplitude of the electromagnetic field and temporal evolutions are simulated by means of the propagation of the field through an optical system. Quantum phenomena can be understood as a consequence of the wave nature of the evolution of quantum states. In this sense, the wave character of the electromagnetic field allows us to simulate, in a pictorical way, the behaviour of the quantum world that usually contradicts common sense. Moreover, the classical electromagnetic field works as an ontologic representation of the wavefunction and it is a useful tool to visualize the structure of problems that are usually complex and counterintuitive.\\ 

Violation tests of Bell inequalities have become a fundamental tool for experimentally proving the presence of entanglement correlations in quantum systems of general interest in areas of quantum information, quantum computation and foundations of quantum mechanics. In a seminal paper \cite{EPR} Einstein, Podolsky and Rosen (EPR) established their argument of the so-called local realism hypothesis.
According to it, if we accept that certain properties of a measured system are present prior to and independent of the observation, then quantum mechanics is not a complete theory of Nature. Almost twenty years later, Bell \cite{Bell} showed that, for systems composed by two spin $1/2$ particles, measurements of some correlated quantities should yield different results in the quantum mechanical case to those expected if we accept the local-realism criterion of EPR. Many experiments confirmed the quantum predictions using Bell-like systems as entangled photons in polarization degrees of freedom \cite{As1, Tit, Zei1, As2, As3, As4, Kwi,GHZ}, entangled photons in transverse momentum degrees of freedom \cite{Paulo} and entangled atoms \cite{Knight}. More recently, a novelty simulation of Bell inequalities violation using RMN techniques was reported \cite{Souza} \\ 

In this paper we will show that an analogy of Bell inequalities violation can be simulated using an optical architecture similar to those used in optical processing. In the Bell experiment, two subsystems share an entangled state of two qbits. In our case the two qbits state is encoded by means of the optical modulation of an electromagnetic wave. The experimental set-up consists in an optical processor with a phase grating in its Fourier plane. The phase grating is represented in a spatial light modulator working in phase mostly mode. The output intensity distribution is registered by a CCD camera and is then analized with an imaging software. We show experimental results for a representation of a entangled state for which a Bell inequality of Clauser-Horne-Shimony-Holt (CHSH) type  is violated \cite{CHSH}. We also discuss the illustrative interpretation of this kind of simulations.\\

The paper is organized as follows: in Section 2 we give a brief review of the basic general concepts of Bell inequalities violation. In Section 3 we present some considerations about optical simulations of quantum information processing by means of optical architectures. We show how quantum states can be represented as images and universal quantum gates can be simulated as coherent optical processors. In section 4, we show how all these elements can be combined in order to obtain the optical setup representation of two separated observers, each one performing measurements of local observables in the Hilbert space of their own qbit. In Section 5 we present experimental results in which a contradiction of the EPR local realism hypothesis can be evidenced. In Section 6 we provide some conclusions and discuss the implicancies of this simulation.\\

\section{Bell inequalities}

We will use in this paper the following notation: a state of the two-dimensional Hilbert space $\textit{H}_{2}$ (or the qbits space) is denoted as a complex linear combination of the two states of the computational basis $\left\{\left|0\right\rangle,\left|1\right\rangle\right\}$. A separable state of the $2^{2}$-dimensional space of a composed system of $2$ qbits $\textit{H}_{2}^{\otimes 2}$ is denoted as the product $\left|\Psi(A)\right\rangle_{A}\otimes\left|\Psi(B)\right\rangle_{B}$, where $\left|\Psi(j)\right\rangle_j= \alpha_{j}\left|0\right\rangle_{j}+\beta_{j}\left|1\right\rangle_{j}$, $j=A,B$; is the quantum state associated to the qbit A or B respectively. In what follows we briefly describe the CHSH approach to Bell type experiments \cite{CHSH}. Let us suppose that two spacelike separated observers (Alice and Bob) share an ensemble of entangled states $\left|\Psi\right\rangle_{AB}=\left(\left|0\right\rangle_{A}\otimes\left|0\right\rangle_{B}+\left|1\right\rangle_{A}\otimes\left|1\right\rangle_{B}\right)/\sqrt{2}$ (subindices $A$ and $B$ denote the observers, or equivalently the qbits they will each measure). Let us consider two pairs of local physical observables: $A$ and $A'$ for Alice; and $B$ and $B'$ for Bob. As usually, we will define these observables as: 

\begin{equation}
A  = \widehat{\alpha}\cdot\vec{\sigma}_{A}, 
\ \ A' = \widehat{\alpha}'\cdot\vec{\sigma}_{A},
\ \ B  = \widehat{\beta}\cdot\vec{\sigma}_{B},
\ \ B' = \widehat{\beta}'\cdot\vec{\sigma}_{B} 
\end{equation}

\noindent where $\widehat{\alpha},\widehat{\alpha}',\widehat{\beta}, \widehat{\beta}'$ are unit vectors and $\vec{\sigma}_{A}$ and  $\vec{\sigma}_{B}$ are vectors whose components are the Pauli matrices operating on the local subspaces associated to Alice and Bob respectively. If Alice and Bob make a random choice of one observable of their pair and perform a simultaneous measurement, they have four possible non local combinations:

\begin{equation}
A \otimes B, \ \ A \otimes B', \ \ A' \otimes B, \ \ A' \otimes B'
\label{dos}
\end{equation}

The measurements of the expected values of such nonlocal quantities can be carried out by performing local measurements. For instance, let us suppose that we want to know the expected value of $A \otimes B$. After finishing a set of many experiments where Alice measures $A$ or $A'$ and Bob measures $B$ or $B'$, they perform the empirical sub-ensemble average of the product of their results $\left\langle A \otimes B \right\rangle=1/N\sum_{i=1}^{N} A_{i}B_{i} $ where Alice has measured A and Bob B. The outcome of each local measurement is $A_{i}, B_{i}=\pm1$ and $N$ is the number of experiments where the Alice's choice was $A$  and Bob's was $B$. It is of most importance to notice that Alice and Bob measure in coincidence so that the measurement which Alice performs does not disturbe the result of Bob's (or vice versa). They can do the same with the three remaining  nonlocal observables $A \otimes B', \ \ A' \otimes B$ and $ A' \otimes B'$. In the end, Alice and Bob will have four independent measurement statistics of four randomly chosen sub-ensembles of the ensemble of singlet states. Alice and Bob can then calculate the expected value of the following nonlocal quantity:

\begin{equation}
O = A \otimes B + A \otimes B'+ A' \otimes B' - A' \otimes B  
\end{equation}

Since Pauli matrices do not commute, quantum mechanics asserts that the four nonlocal observables of eq.[\ref{dos}] are not compatible and therefore cannot be simultaneously measured. Moreover, the Bohr principle of complementarity \cite{Bohr} claims that we are forbidden to consider simultaneously the possible outcomes of mutually exclusive experiments. However, according to the local realism hypothesis, there exist local hidden parameters which completely determine the outcomes of the chosen measurement, having predictions quite different to those predicted by quantum mechanics. Moreover, these parameters also determine the outcomes that we would have if we have measured an observable  which is incompatible with that actually measured. We have no control on hidden parameters and so there are some degrees of freedom that are not precisely known. In the case of the Bell experiment, each hidden parameter assigns well defined outcomes $\pm1$ to each local measurement. It has been demonstrated that under such condition the expected value of the nonlocal quantity of Eq.[3] satisfies the CHSH form of the Bell inequality \cite{CHSH}:

\begin{equation}
\left|\left\langle O \right\rangle\right| = \left|\left\langle A \otimes B\right\rangle +  \left\langle A \otimes B'\right\rangle +\left\langle  A' \otimes B'\right\rangle- \left\langle  A' \otimes B \right\rangle \right|\leq 2
\end{equation}

As mentioned in the introduction, quantum experiments measuring different observables have been performed \cite{As1, Tit, Zei1, As2, As3, As4, Kwi,GHZ}. We will inspect the particular case $\widehat{\alpha'}=\widehat{\beta'}= \widehat{z}$ so that $A' = B'\equiv C$. With this assumption, $\left\langle A'\otimes B'\right\rangle= \left\langle C\otimes C\right\rangle=\widehat{z}\cdot\widehat{z}=+1$ for the state we have chosen, and the left Bell inequality [4] becomes:

\begin{equation}
\left\langle O \right\rangle = \left\langle A \otimes B\right\rangle + \left\langle A \otimes C\right\rangle - \left\langle  C\otimes B \right\rangle \leq 1
\end{equation}

The equation above must be satisfied for arbitrarly chosen observables $A$ and $B$ with $C=\sigma_z$. However, without loose of generality and for practical reasons, we will set observables whose optical implementation is relatively simple as it will be clear in section 3. These observables are: 

\begin{equation}
A=\sigma_{x}, \ \ B = \sin{\theta}\sigma_{x}+cos{\theta}\sigma_{z}, \ \ C = \sigma_{z} 
\end{equation}

Optical simulations of measurements of $\sigma_{x}$ and $\sigma_{z}$ by means of imaging proccesing architectures have been implemented in previous works \cite{FranHad, FranQRW, FranTel}. We will show in the next section that measurement of $\sin{\theta}\sigma_{x}+cos{\theta}\sigma_{z}$ can be performed in a similar way by controlling additional experimental parameters.\\

Additionally, we will emphasize the rol of two-particle interference effects in the mechanism of Bell inequalities violation in this classical analogy. In order to make the rol of such interference effects to become apparent, we will analyze both the  case of maximally entangled states and the case of mixed states. Let us consider the expectation value of the quantity $O$ with respect to the mixed state whose density matrix is the convex sum of the pure entangled state $\rho_{pure}=\left|\Psi\right\rangle_{AB}\left\langle \Psi\right|_{AB}$ and the maximally mixed state: $\rho_{mixed}=\frac{1}{2}\left[\left(\left|0\right\rangle_{A}\otimes\left|0\right\rangle_{B}\right)\left(\left\langle 0\right|_{A}\otimes\left\langle 0\right|_{B}\right)+\left(\left|1\right\rangle_{A}\otimes\left|1\right\rangle_{B}\right)\left(\left\langle 1\right|_{A}\otimes\left\langle 1\right|_{B}\right)\right]$. Therefore the state can be written as:

\begin{equation}
\rho=q\rho_{pure} +(1-q)\rho_{mixed}=\frac{1}{2}\left(\begin{array}{cccc}
                                              1&0&0&q\\
                                              0&0&0&0\\
                                              0&0&0&0\\
                                              q&0&0&1\end{array}\right); 
\end{equation}                                                                                                                                    
                                                                                                                                                                                                                                                    where the matrix representation is given in the computational basis $\left\{\left|0\right\rangle_{A},\left|1\right\rangle_{A}\right\} \otimes \left\{\left|0\right\rangle_{B},\left|1\right\rangle_{B}\right\}$ and $q\in\left[0,1\right]$.  The state of eq. [7] is a pure state for $q=1$ and becomes mixed if $0\leq q < 1$ owing to the lost of the off-diagonal coherence components of the density matrix.  The maximally mixed state corresponds to $q=0$. Evaluating  the expected value of the quantity $O$  with respect to the state [7] and setting the observables defined in [6]  we have:
                                                                                                                                                                                                                                                                                         
\begin{equation}
\left\langle O(\theta)\right\rangle = tr\left[\left(A \otimes B\right)\rho\right]+ tr\left[\left(A \otimes C\right)\rho\right]  - tr\left[\left(C\otimes B\right)\rho\right] = q \sin{\theta}-\cos{\theta}
\end{equation}   

This last result contradicts the Bell inequality [8] if $\theta_{c}\left(q\right)<\theta< \pi$ where $\theta_{c}\left(q\right)$ is the solution of the equation $q=\left(1+\cos{\theta}\right)/\sin{\theta}$ that always exists in the interval $0 <\theta< \pi$ for all $0<q\leq1$. Maximal violation occurs for $q=1$ and corresponds to the Bell state which is pure and maximally entangled. On the contrary, there is no quantum correlations in the case of the maximally mixed state $q=0$ and therefore the Bell inequality is not violated. In the next section we will describe the problem underlined above with elements of classical optics.\\                                                                                                                                                                                                                                             
\section{Optical simulation}

In order to perform the optical simulation of the Bell experiment, we will present first the way of encoding the complex amplitudes of the qbits as spatial distributions of light and we will emulate the unitary evolutions by means of an optical system composed of common optical devices such as lenses, phase gratings and phase shifters. These devices modify the complex amplitude of the electromagnetic field arising from the input image, where the state is encoded, reproducing the temporal evolution of the quantum state. The final state is obtained as an image that can be registered by a CCD. The spatial distribution of the intensity generated  by the final complex amplitude can be interpreted as the probability distribution of finding a given state.\\ 

\subsection{Encoding qbits in optical scenes}

We use a method which is extensively discussed in the literature \cite{Puen, Spr98, Spr01} based on the representation of qbits as position C-bits  and constitutes one possible scheme of the so called "unary representation of quantum systems". 
According to it, in the input scene we encode the logical values $\left|0\right\rangle$ and $\left|1\right\rangle$ of a single qbit in two slices located in the left and right halves of the full plane respectively. Let $(x_{o},y_{o})$ be the coordinates of the input plane. As the representation is one-dimensional, we have traslational symmetry with respect to the $y_{o}$ axis, and we can consider only the $x_{o}$ coordinate. According to wavefunction formalism, the optical analogy suggests the following notation:

\begin{eqnarray}
\left\langle x_{o}\right.|0\rangle & \equiv & Rect\left(\frac{x_{o}+a}{b}\right)  \nonumber \\
\left\langle x_{o}\right.|1\rangle & \equiv & Rect\left(\frac{x_{o}-a}{b}\right)
\end{eqnarray}

\noindent where $Rect(x)$ is a unit rectangle function that takes the value $1$ if $\left|x\right|\leq 1/2$ and $0$ in other case. Eq. [9] describes unit amplitude transmittance in a rectangle of width $b$ centered in $x=\pm a$ ($b\ll a$) where the computational state $\left|0\right\rangle$ is associated to the left rectangle and the computational state $\left|1\right\rangle$ is associated to the right rectangle as it is shown in Fig.1(a). To encode the more general state of two qbits, modulation of the complex amplitude of the field in both slices is needed as we can see in Fig.1(b). The optical analogy of the quantum measurement process is very simple. In fact, if the state $\left|\varphi\right\rangle=\alpha\left|0\right\rangle+\beta\left|1\right\rangle$  is represented, the relative intensities of the slices will be precisely $\left|\alpha\right|^{2}$ and $\left|\beta\right|^{2}$ and, after renormalization $\left|\alpha\right|^{2}+\left|\beta\right|^{2}=1$, they can be interpreted as the probabilities associated to the measurement outcomes $+1$  and $-1$ in a projective measurement in the computational basis. Making a step forward in this classical analogy, a quantum measurement is equivalent to keeping the left slice with probability $\left|\alpha\right|^{2}$ or the right slice with probability $\left|\beta\right|^{2}$. Moreover, the expected value of $\sigma_{z}$ on the state represented, can be calculated by extracting information from the image represented in Fig. [1.b] as $\left\langle \sigma_{z}\right\rangle=tr(\sigma_{z}\left|\varphi\right\rangle\left\langle \varphi\right|)=(\left|\alpha\right|^{2}-\left|\beta\right|^{2})/(\left|\alpha\right|^{2}+\left|\beta\right|^{2})$.

\begin{figure}[ht]
\centering
\includegraphics [scale = 1]{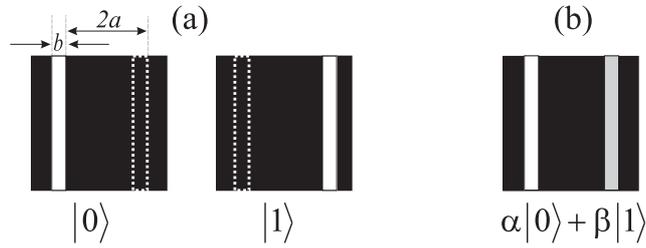}
\caption{\small{Optical representation of the single qbit state. (a) Optical representation of the states of the computational basis. (b) The input scene associated to the optical single qbit. The state $\alpha\left|0\right\rangle+\beta\left|1\right\rangle$ is represented by the two single "left" or "right" slices where the  constants $\alpha$ and $\beta$ are the complex amplitudes of the electromagnetic field in each computational slice}}
\end{figure}

The scheme described above is easily generalizable to represent two or more qbits. In order to represent the computational states of a second qbit we will define another dicotomic regions within each one of the previously defined regions. As we can see in Fig.2, we will use the convention of splitting the plane in up and down regions to accomplish this. We denote $A$ the qbit encoded in the up-down direction and $B$ the qbit encoded in the left-right direction. In this case the representation of the two qbits basis will be a two dimensional extension of Eqs.[9]. 

\begin{figure}[ht]
\centering
\includegraphics [scale = 1 ]{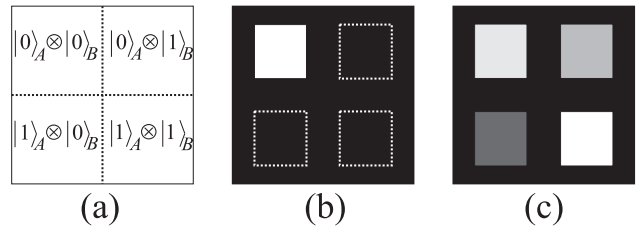}
\caption{\small{Schematic picture of the representation of two qbits states by using optical scenes. (a) Spatial organization of the input plane in order to emulate two qbits states. (b) Optical representation of the $\left|0\right\rangle_{A}\otimes\left|0\right\rangle_{B}$ state. (c) Optical representation of the general pure two qbits state. Gray level scale correspond to different amplitudes and phase modulations of the classical wavefront }}
\label{fig:Bell1}
\end{figure}

\subsection{Pure and mixed states: the optical approach}

In the previous section we have discussed the optical representation and the classical interpretation of the measurement process in the case of pure quantum states. We will briefly introduce here how we can extend these concepts in the case of mixed states. A mixed state with density matrix $\rho=\sum_{i}p_{i}\left|\varphi_{i}\right\rangle\left\langle \varphi_{i}\right|$; $p_{i}\geq0$, $\sum_{i}p_{i}=1$, consists in a set of pure states $\left\{\left|\varphi_{i}\right\rangle, i=1,2,....,n\right\}$, each appearing with its respective probability $p_{i}$.
The strategy we use to simulate and measure mixed states is as follows: first we represent an image in the input plane that simulate the member of the ensemble $\left|\varphi_{i}\right\rangle$, which is a pure state. Then we calculate the expected value of $\sigma_{z}$, $\left\langle \sigma_{z}\right\rangle$,  as we have discussed in section 3.1, then we multiply the result by $p_{i}$,  and finally we sum over all the results. In this way, since $tr(\sigma_{z}\rho)=\sum_{i}p_{i}tr(\sigma_{z}\left|\varphi_{i}\right\rangle\left\langle \varphi_{i}\right|)$ the same values as having a statistical ensemble, will be obtained.
It is worth mentioning that following this procedure another analogy can be achieved using a temporal succession of images with a duration $T_i$. If the total time of the experiment is $T$, then we select the times so that $T_{i}/T=p_{i}$. In this case, one has to integrate the output images in the time T of the experiment. This could be realized using an optical element capable of modulating the light field in the input plane in a dynamical way. For practical reasons, we used the first strategy throghout our work.

\subsection{The optical U(2) operator}

The key of our representation is the possibility of simulate the unitary change of basis that maps the complex amplitudes of the input state in the computational basis onto the complex amplitudes of the state in the basis of eigenstates of the observables that should be measured. This simulation is accomplished by using phase gratings as filters in an optical processor architecture. In what follows we denote the field amplitude as depending on a single relevant coordinate due to the unidimensional character of the optical simulation of local $U(2)$ operators. The simulation of U(2) operators acting on single qbit states works as follows. The quantum state is encoded in an input scene located in the previous focal plane of a spherical lens of focal distance $f$. For collimated illumination, this lens allows to obtain on its back focal plane the Fourier transform of the input scene, which corresponds to a field distribution whose relevant coordinate is $x_{F}$. The relationship between the spatial frequency variable  $f_{x}$ and the position coordinate $x_{F}$ on the Fourier plane  is $f_{x}=x_{F}/\lambda f$, where $\lambda$ is the wavelength of the light field. In the Fourier plane, a spatial filter of complex transmittance $H\left(f_{x}\right)$ is placed. 
A second spherical lens of focal distance $f$ is placed so that its previous focal plane lies in the Fourier plane. This lens allows to obtain the inverse Fourier transform of the product between the Fourier transform of the input transmitance and the function $H\left(f_{x}\right)$. The field amplitude in the output plane is the convolution between the input complex amplitude and the so called impulse response of the system that is defined as the inverse Fourier transform of the function $H\left(f_{x}\right)$ \cite{Good}. In our case, spatial filtering in the Fourier plane is performed by an almenary phase grating which is a square wave phase modulation of amplitude $0<\phi<2\pi$ and spatial period $2p$. We denote the width of each square pulse as $p$ and the position of the center of the pulse in the frequency domain $f_{c}$. The complex transmittance of the filter is:

\begin{equation} 
H\left(f_{x}\right) = \left\{ \begin{array}{ll}
e^{i\phi} & \textrm{if $\left|f_{x}-f_{c}\right| < p/2$}\\
1 & \textrm{in other case}
\end{array} \right.
\end{equation}

\noindent where the function above is defined in $f_{x}\in \left[-p,p\right]$ and is extended by periodicity for all $f_{x}\in \Re$. The complex transmitance defined in Eq. [10] could be expanded in terms of pure harmonic components so that the function $H(f_{x})$ is written as $H(f_{x})=\sum_{n\in\ Z}C_{n}\exp\left(i\frac{\pi n}{p}f_{x}\right)$. Taking into account only the three central diffracted orders (coefficients $C_{0}$ and $C_{\pm1}$), the  inverse Fourier transform of the equation above is: 

{\setlength\arraycolsep{1pt}
\begin{eqnarray}
h(x) & = & \int_{-\infty}^{+\infty}H(f_{x})\exp{(i 2\pi f_{x}x)}df_{x} \nonumber \\
                                   & = & \cos\frac{\phi}{2}\delta(x)+\frac{2}{\pi}\sin\frac{\phi}{2} e^{i\gamma^{-}}\delta\left(x+\frac{1}{2p}\right) +
                                   \frac{2}{\pi}\sin\frac{\phi}{2} e^{i\gamma^{+}}\delta\left(x-\frac{1}{2p}\right);\nonumber \\
                                   \end{eqnarray}}

where we have defined the phase constants $\gamma^{\pm}=\frac{\pi}{2} \pm \frac{\pi}{p}f_{c}$. Let us suppose now that in the input scene we represent a complex linear combination of the two computational states $\left\langle x_{o}\right|\left.\Psi_{in}\right\rangle=\alpha\left\langle x_{o}\right|\left.0\right\rangle+\beta\left\langle x_{o}\right|\left.1\right\rangle$ according to Eq.[9]. The output signal $\left\langle x_{i}\right|\left.\Psi_{out}\right\rangle$ will be equal to $\left(h \ast \Psi_{in}\right)_{(x_{i})}$ corresponding to the convolution between the input signal and the impulse response of the system defined in Eq. [11] evaluated in the coordinate $x_{i}$ of the image plane. The result gives six terms that are the three principal diffracted orders of the two slices that represent the computational states. Under certain conditions that can be experimentally controlled, the expression can be simplified. In fact, if we choose a grating whose spatial frequency satisfies the relationship $2p=1/2a$ with respect to the spatial separation $2a$ of the slices, the separation of the diffracted orders in the final plane will be equal to the distance of the two slices, allowing the interference between them, as it is suggested in Fig.[3]. All diffracted orders located out of the computational regions are not registered.\\

The input-output relation of the process in matrix form, using the identifications of Eq.[9], is expressed in the representation of the computational basis as:

\begin{equation}
\left(\begin{array} {c} \alpha \\
                        \beta
      \end{array}\right)\longrightarrow \left(\begin{array} {c} \alpha' \\
                        \beta'
      \end{array}\right)=\left(\begin{array}{cc}\cos\frac{\phi}{2}&
     \frac{2}{\pi}\sin\frac{\phi}{2} e^{i\gamma^{-}}\\ \frac{2}{\pi}\sin\frac{\phi}{2} e^{i\gamma^{+}} & \cos\frac{\phi}{2}\\\end{array}\right)\left(\begin{array} {c} \alpha \\
                        \beta
      \end{array}\right)
      \end{equation}
\\

where $\phi$ and $\gamma^{\pm}$ are real-valued. The process is schematized in Fig.[3]. The similarity between the general expresion of U(2) operators and the $2$-parameter family of linear operators of Eq. [12] becomes evident. For instance, if we want to measure the observable $\sigma_{x}$, we must apply the Hadamard operator $H$ \cite{Nie}, and then perform the measurement in the computational basis. This can be done by setting $\tan\left(\phi/2\right)=2/\pi$ and $f_{c}=-p/2$. The Hadamard-like operator $\sqrt{2}H \sigma_{z}$ obtained in this way, can be transformed in the proper Hadamard operator by placing a phase plate $\sigma_{z}$ in the front of the first lens since $\sigma_{i}^{2}=1$ $\forall i$ after renormalization. Measurement of hermitian observables defined in the Hilbert space of the two level quantum system can be simulated in the same way. In the Table [1] we show the unitary change of basis associated to each observable of Eqs.[6], and the corresponding values of the parameters $\phi$ and $f_{c}$. A phase shift $\sigma_{z}$ in front of the first lens must be eventually included.

\begin{table}[h!]
	\centering
		\begin{tabular}{ccccc}
		$Hermitian$                                    & $Unitary$                                & $\phi$                           & $f_{c}$  & $phase$ \\
		\hline
		$\sigma_{x}$                                   & $H$                                      & $2Arctan\left(2/\pi\right)$      & $-p/2$   &  $\sigma_{z}$\\
		$\sin{\theta}\sigma_{x}+cos{\theta}\sigma_{z}$ & $e^{-i\theta\sigma_{y}/2}\sigma_{z}$ & $2Arctan\left((2/\pi) Tan\left(\theta/2\right)\right)$      & $-p/2$ & $\sigma_{z}$ \\
		$\sigma_{z}$                                   & $1$                                      &  -                               & -      & -\\
		\hline 
		\end{tabular}
		\caption{Hermitian observables, unitary change of basis and parameters of the optical simulation}
\end{table}

\begin{figure}[ht]
\centering
\includegraphics [scale = 1]{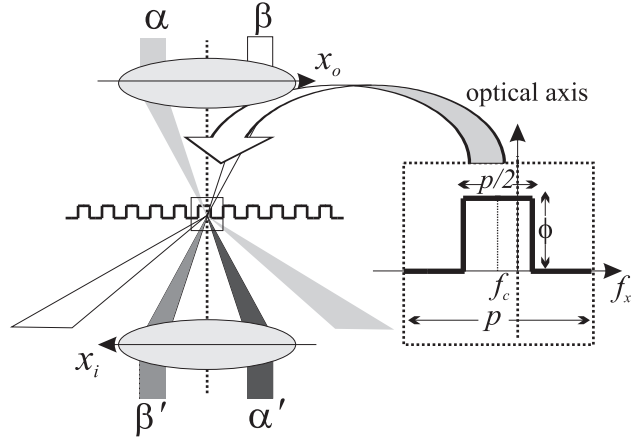}
\caption{\small{Schematic picture of the optically simulated U(2) operation. Complex amplitudes $\alpha$ and $\beta$ are mapping onto  $\alpha'$ and $\beta'$ by means of a 4f coherent optical processor with an almenary phase grating in the Fourier plane.}}
\end{figure}

\section{Optical implementation of the analogy} 

The experiment reported in this section, works as a classical optics analogy of the Bell experiment. We will test our setup in two cases which correspond to the two states $q=0$ and $q=1$ described in Eq.[7]:

\begin{itemize}
	\item \textbf{Case 1:} $q=1$. Alice and Bob will share an entangled pair. For simulating this, we encode the maximally entangled state  $\left|\Psi\right\rangle_{AB}=\left(\left|0\right\rangle_{A}\otimes\left|0\right\rangle_{B}+\left|1\right\rangle_{A}\otimes\left|1\right\rangle_{B}\right)/\sqrt{2}$ by uniformly illumination of the top-left \emph{and} the down-right quarters of the full input plane (Fig 4.a).
	\item \textbf{Case 2:} $q=0$. The input state will be the statistical mixture represented by the density matrix $\rho_{mixed}=\frac{1}{2}\left[\left(\left|0\right\rangle_{A}\otimes\left|0\right\rangle_{B}\right)\left(\left\langle 0\right|_{A}\otimes\left\langle 0\right|_{B}\right)+\left(\left|1\right\rangle_{A}\otimes\left|1\right\rangle_{B}\right)\left(\left\langle 1\right|_{A}\otimes\left\langle 1\right|_{B}\right)\right]$ whose optical analogy is the uniform illumination of the top-left \emph{or} the down-right quarters of the full input plane, each with probability $1/2$. This is an incoherent superposition of the computational states and it should not have any quantum correlation (Fig 4.b). 
	\end{itemize}
	
	\begin{figure}[ht]
\centering
\includegraphics [scale = 1 ]{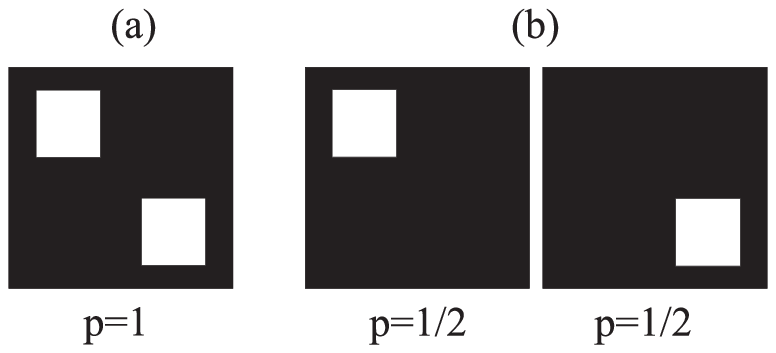}
\caption{\small{Optical representation of the maximally entangled state $\left(\left|0\right\rangle_{A}\otimes\left|0\right\rangle_{B}+\left|1\right\rangle_{A}\otimes\left|1\right\rangle_{B}\right)/\sqrt{2}$ (a) and of the maximally mixed state $\frac{1}{2}\left[\left(\left|0\right\rangle_{A}\otimes\left|0\right\rangle_{B}\right)\left(\left\langle 0\right|_{A}\otimes\left\langle 0\right|_{B}\right)+\left(\left|1\right\rangle_{A}\otimes\left|1\right\rangle_{B}\right)\left(\left\langle 1\right|_{A}\otimes\left\langle 1\right|_{B}\right)\right]$} (b) as optical scenes.}
\end{figure}

The complete optical set-up is schematized in Fig.[5]. An Argon laser source ($\lambda= 477 nm$) is filtered and then collimated with lens $L_{0}$. The collimated beam impinges onto the binary mask $P_{i}$ which represents the two qbit state that Alice and Bob use during the experiment. The input scene is placed in the previous focal plane of the lens $L_{1}$ (focal length 26 cm) that allows to obtain the Fourier transform of the input in its back focal plane, or Fourier plane.  In the Fourier plane, we place the spatial filter for simulating two local unitary operators. According to previous discussions, this can be done with the composition of two orthogonal almenary phase gratings as showed in Fig. [6]. Horizontal phase grating with phase modulation parameter $\phi_{A}$ produce diffracted orders in the "up-down" direction in where Alice's qbit is encoded. Vertical phase grating with parameter $\phi_{B}$ produce diffracted orders in the "left-right" direction associated to Bob's subsystem. The two dimensional almenary phase grating, whose phase modulation goes from $0$ to $\phi_{A}+\phi_{B}$ $(mod 2\pi)$, was programmed in a spatial light modulator (SLM). This device consists in a Sony liquid crystal display TV (LCTV) that combined with two polarizers  ($P_{1}$ and $P_{2}$) and two quarter wave plates ($QWP_{1}$ and $QWP_{2}$), acts as a mostly phase modulator \cite{Clau}. The LCTV (model LCX012BL) was extracted from a commercial video-projector and is a VGA resolution panel (640 x 480 pixels) with square pixels of 34 $\mu$m size separated by a distance of 41.3 $\mu$m. The process is completed with the lens $L_{2}$ (focal length 26 cm) , that allows to obtain the inverse Fourier transform. The final image in the output plane $P_{o}$ is captured by a videocamera (CCD). 

\begin{figure}[ht]
\centering
\includegraphics [scale = 1 ]{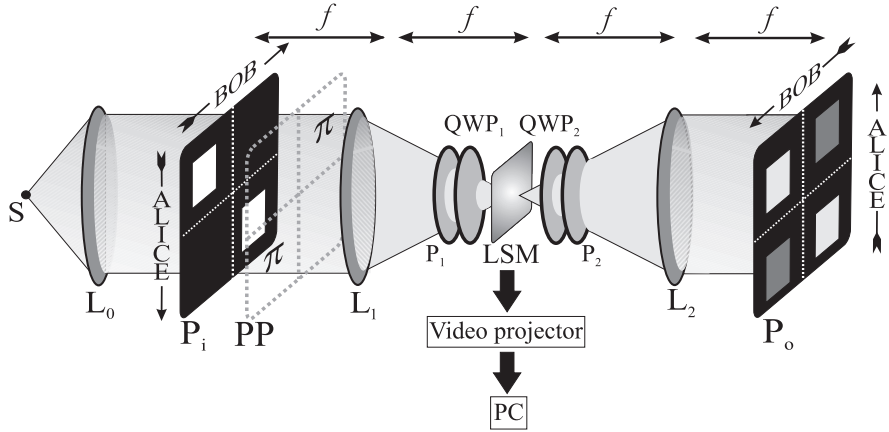}
\caption{\small{Experimental setup for simulating Bell experiment as an imaging system.}}
\end{figure}

According to Table [1], an adittional phase plate (PP) that introduces a $\pi$ phase shift in the left-bottom and in the right-top quarters of the input wavefront must be included. Therefore, this phase plate does not affect the illuminated zone of the input plane and we can ignore it. Moreover, in the final image, we must take into account the  inversion of the coordinates system whose senses are indicated with arrows on the $P_{i}$ and  $P_{o}$ input and output planes respectively in Fig.[5]. The protocol of the full experiment is depicted in Fig.[6] and can be described as follows:

\begin{figure}[ht]
\centering
\includegraphics [scale = 1]{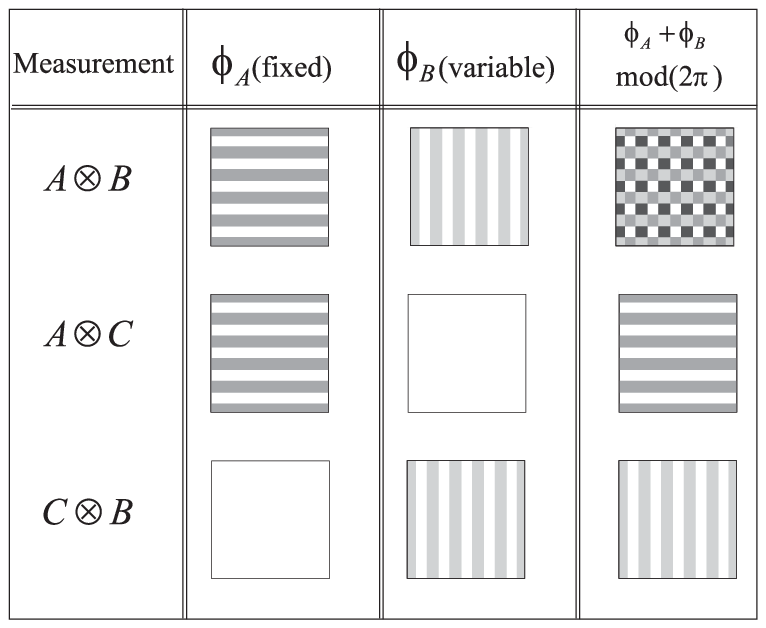}
\caption{\small{Detail of the protocol of the full experiment.}}
\end{figure}

Each local operation is simulated by using one of the two orthogonal almenary phase gratings. Horizontal modulation from $0$ to $\phi_{A}$ simulate unitary operations on the Alice "up-down" encoded qbit while vertical modulation from $0$ to $\phi_{B}$ works equally for the Bob "left-right" encoded qbit. Alice measurement is fixed to $\sigma_{x}$  (or $\sigma_{z}$)  and therefore, eventually she only applies a Hadamard operator before projecting her qbit on the computational basis. Bob measurement is varying as $\sin{\theta}\sigma_{x}+cos{\theta}\sigma_{z}$ for $0\leq \theta\leq2\pi$ and the phase modulation $\phi_{B}(\theta)=2Arctan\left((2/\pi) Tan\left(\theta/2\right)\right)$ (see Table [1]) has to be variable. In both cases the amplitude of the phase modulation is controlled directly by the spatial light modulator. Measurement of $\sigma_{z}$ is performed by orthogonal projection on the computational basis and no phase modulation is needed (the white squares in Fig. [7]). The output scenes corresponding to the three pairs of local measurements (one fixed $A\otimes C$ and two varying $A \otimes B$ and $C \otimes B$) are registered by the CCD and recorded for its posterior analysis. The analysis method and the corresponding results will be shown in the next section.

\section{Bell inequalities: experimental results}

Once the output scenes corresponding to each measurement are obtained, the mean value of the observable $\left\langle O\right\rangle$ of Eq.[8] can be easyly evaluated. In fact, the output image contains all the information about the statistical properties of the measurement for a two qbits system. Unitary changes of basis reduce the problem of the measurement of an arbitrary observable into the problem of measuring $\sigma_{z} \otimes \sigma_{z}$. This is a projective measurement in the computational basis $\left\{\left|0\right\rangle_{A},\left|1\right\rangle_{A}\right\} \otimes \left\{\left|0\right\rangle_{B},\left|1\right\rangle_{B}\right\}$. The results of a projective measurement in computational basis can be easy interpreted in terms of the distributed intensities of the output field obtained in our experiment. The underlying process in the analysis of the output images is depicted in Fig. [7]. In what follows $P(A_{i},B_{j})$ with $i,j=\pm1$ is the joint probability of, after a projective measurement in computational basis, $A=i$ and $B=j$; i.e. the measurement outcomes obtained by Alice and Bob were $i$ and $j$ respectively. According to elementary probability theory, this quantity can be evaluated as $P(A_{i},B_{j})= P(B_{j}/A_{i})P(A_{i})$; where $P(B_{j}/A_{i})$ is the conditional probability of $B=j$ knowing that the result $A=i$ was obtained, and $P(A_{i})$ is the probability that the result $A=i$ was obtained, independently of Bob.\\

\begin{figure}[ht]
\centering
\includegraphics [scale = 1 ]{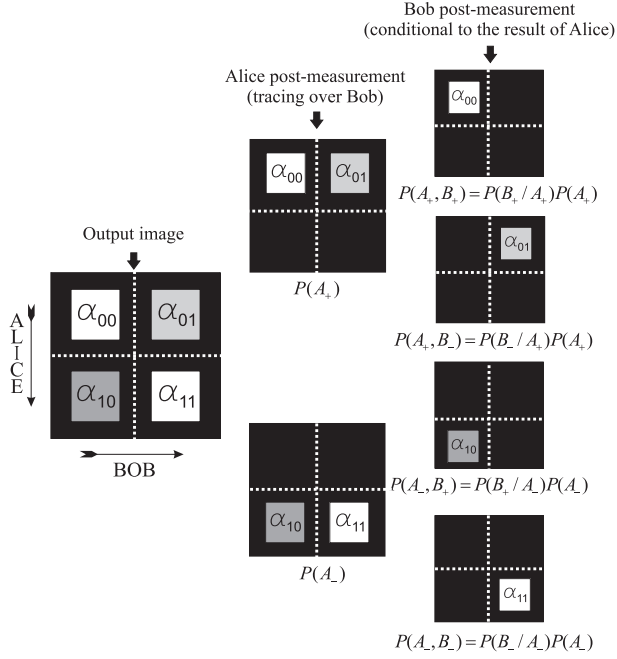}
\caption{\small{Measuring a two qbits system in computational basis from the output distribution intensities.}}
\end{figure}

Let us suposse that Alice performs the measurement of her qbit. As we have discussed above, it means that she has to keep the up or the down half of the full plane with different probabilities. The statistical properties of Alice's measurement are described by the marginal probabilities deduced from the joint distribution integrated over the degree of freedom associated to Bob. In the formalism of quantum mechanics, marginal distribution of Alice is defined by the reduced density matrix $tr_{B}(\rho_{AB})$. From the optical point of view, partial trace over Bob's subsystem means that the accesible information available for Alice is related to the field distribution in the "up-down" direction, independently to the field distribution in the "left-right" direction in where the information available for Bob is encoded. So, the two possible post-measurement states of the full system with their respective probabilities will be:\\

\begin{equation} \left|0\right\rangle_{A}\otimes\left(\alpha_{00}\left|0\right\rangle_{B}+\alpha_{01}\left|1\right\rangle_{B}\right)/\sqrt{\left|\alpha_{00}\right|^2+\left|\alpha_{01}\right|^2}
\end{equation}
\\

with probability $P(A_{+})=(\left|\alpha_{00}\right|^2+\left|\alpha_{01}\right|^2)/(\left|\alpha_{00}\right|^2+\left|\alpha_{01}\right|^2+\left|\alpha_{10}\right|^2+\left|\alpha_{11}\right|^2)$, that means that Alice measurement outcome was $+1$ and her pseudorandom choice was the top half of the scene; or:\\

\begin{equation} \left|1\right\rangle_{A}\otimes\left(\alpha_{10}\left|0\right\rangle_{B}+\alpha_{11}\left|1\right\rangle_{B}\right)/\sqrt{\left|\alpha_{10}\right|^2+\left|\alpha_{11}\right|^2} 
\end{equation}
\\
with probability $P(A_{-})=(\left|\alpha_{10}\right|^2+\left|\alpha_{11}\right|^2)/(\left|\alpha_{00}\right|^2+\left|\alpha_{01}\right|^2+\left|\alpha_{10}\right|^2+\left|\alpha_{11}\right|^2)$, that means that the outcome was $-1$ and her choice was the bootom half of the scene.\\

Now is the turn of Bob. The post-measurement state of the full system after Bob measurement will be conditioned for the result previously obtained by Alice. The four possible post-measurement states are naturally the four computational states. For instance, the post-measurement state $\left|0\right\rangle_{A}\otimes\left|0\right\rangle_{B}$ can be obtained with probability $P(B_{+}/A_{+})=\left|\alpha_{00}\right|^2/(\left|\alpha_{00}\right|^2+\left|\alpha_{01}\right|^2)$ with the previous knowledge that the post-measurement state obtained by Alice was the one described in Eq.[13]. At this point the non-local aspects of the joint measurement become evident. In this last case, the full process is a joint projective measurement of the output state in computational basis in where the result $A=+1$, $B=+1$ is obtained. The corresponding post-measurement state will be  $\left|0\right\rangle_{A}\otimes\left|0\right\rangle_{B}$ with probability $P(A_{+},B_{+})=P(B_{+}/A_{+})P(A_{+})=\left|\alpha_{00}\right|^2/(\left|\alpha_{00}\right|^2+\left|\alpha_{01}\right|^2+\left|\alpha_{10}\right|^2+\left|\alpha_{11}\right|^2)$. The remaining three computational states, can be equally obtained as post-measurements states with probabilities depending on the intensity distribution $\left|\alpha_{mn}\right|^2$; $m,n=0,1$. Therefore, the expected value of $\sigma_{z} \otimes \sigma_{z}$ in terms of the output intensity distribution is:

\begin{eqnarray}
\left\langle \sigma_{z} \otimes \sigma_{z} \right\rangle
								&=&
	P(A_{+},B_{+})+P(A_{-},B_{-})-	P(A_{+},B_{-})-P(A_{-},B_{+}) \nonumber\\
 &=&\frac{\left|\alpha_{00}\right|^2+\left|\alpha_{11}\right|^2-\left|\alpha_{01}\right|^2-\left|\alpha_{10}\right|^2}{\left|\alpha_{00}\right|^2+\left|\alpha_{01}\right|^2+\left|\alpha_{10}\right|^2+\left|\alpha_{11}\right|^2} 
 	\end{eqnarray}

It should be pointed out that Alice and Bob measurements involve only local operations. So the result of the previous analisys does not depend on the chronological order of the measurements since local operators commutes with each other. This is important since Alice and Bob measure simultaneously according to the EPR locality hypothesis.\\

\begin{figure}[ht]
\centering
\includegraphics [scale = 1]{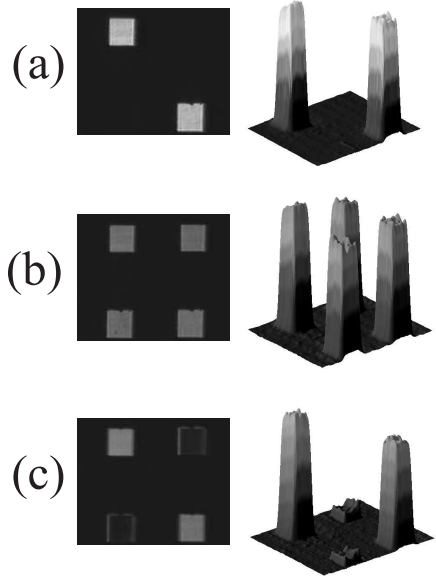}
\caption{\small{First column: some typical output images. Second column: 3D plotting of intensity profiles. (a) Input state. (b) Output state after measurement of $\sigma_{z}\otimes\sigma_{x}$. (c) Output state after measurement of $\sigma_{x}\otimes\sigma_{x}$.}}
\end{figure}	

Calculations involved in Eq.[15] are performed directly from the relative output intensities associated to the computational states in the output image. Typically, images and 3D intensity profiles such as shown in Fig.[8] are obtained. The experiment consists in performing a sampling on the phase modulation of the vertical grating with $0\leq\phi_{B} < 2\pi$, and for each value of $\phi_{B}$ evaluate the quantity $\left\langle O \right\rangle = \left\langle A \otimes B\right\rangle + \left\langle A \otimes C\right\rangle - \left\langle  C\otimes B \right\rangle$ in function of $\theta= 2 Arctan\left((\pi/2) tan\left(\phi_{B}/2\right)\right)$ by inspecting the output images with the observables $A$, $B$ and $C$ defined as in Eq. [9]. The experimental results for $\left\langle O \right\rangle$ vs. $\theta$ are plotted toghether with the theoretical expected result $q \sin{\theta}-\cos{\theta}$. Theoretical and experimental curves are compared with $1$ in order to explore possible violations of the inequality  $q \sin{\theta}-\cos{\theta}\leq 1$. Experimental results are summarized in Fig. [9].\\

\begin{figure}[ht]
\centering
\includegraphics [scale = 1 ]{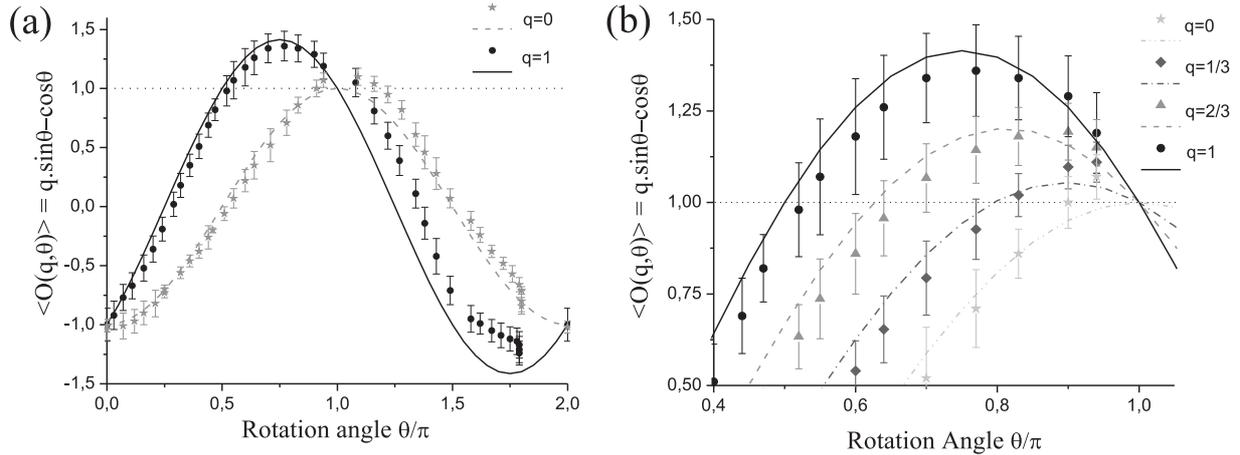}
\caption{\small{Experimental results. (a) Theoretical and experimental $\left\langle O \right\rangle$ vs. $\theta$ plotted in the full range $ \theta \in [0,2\pi)$ for the maximally entangled state $q=1$ and for the mixed state $q=0$. (b) Theoretical and experimental results corresponding to mixed states with density matrix defined in Eq.[7] for $q=1$, $q=2/3$, $q=1/3$ and $q=0$ in the Bell inequality violation domain.}}
\end{figure}

In Fig.[9.a] $\left\langle O \right\rangle$ vs. $\theta$ is plotted in the full range $ \theta \in [0,2\pi)$ for the maximally entangled state $q=1$ and for the mixed state $q=0$. As we can appreciate, although the experimental points differ slightly from the theoretical curves, a good cualitative agreement is obtained. It has to be mentioned that the  SLM was used in maximal resolution of two pixels per spatial period of the grating. Since the properties of phase modulation of the SLM in high resolution are far from optimal, some significatively differences between experimental points and the theoretical curve appear, mainly in the range $ \theta \in [\pi,2\pi)$ of Fig.[9.a]. Such differences are more significant in the case $q=1$ in where they are amplified by interference effects. Maximal violation of the Bell inequality occurs for $q=1$ and there is no violation for $q=0$ as expected.  In Fig.[9.b] experimental and theoretical results corresponding to mixed states with density matrix defined in Eq.[7] are showed for $q=1$, $q=2/3$, $q=1/3$ and $q=0$. In this case the plot is shown not in the full range but in the zone of violation of the Bell inequality. The experimental results of the simulation are also in good agreement with the theory, within the experimental error. The lost of coherence of the state, when the parameter goes from the maximally entangled to the maximally mixed state becomes evident in the transition from maximal violation to no violation.\\

\section{Conclusions}

We have implemented an optical setup to classically simulate Bell inequalities violation. We have shown how a conventional optical processing architecture can be used to optically simulate a Bell-type experiment scenario in where two space-like separated observers  evaluate certain statistical correlations of non local observables by means of tensor products of local measurements. We have optically simulated all the elements corresponding to  the real quantum process showing that all these elements have its classical optic counterpart. We have obtained experimental results by using the optical proposed set up. The simulation begins with the optical representation of the quantum state of a four-dimensional Hilbert space as an image organized in four quarters according to the statement in Sec. 3.1. In this unary representation and according to Sec. 3.2, pure and mixed states can be emulated. The encoded information is organized as a bipartite two level system. One part, conventionally called "Alice" has  available dicotomic information related to the field distribution in the "up-down" direction of the scene. The second, "Bob" is associated to the ortogonal "left-right" direction. Quantum non locality arises from the assumption that the information available for one of the observers is unavailable for the other and vice versa. Then, we process the input image with an optical set-up composed by two sperical lenses and a bi-dimensional phase grating between them. The proposed architecture is in essential a 4f coherent optical processor with a phase grating in the Fourier plane.
The optical processor is designed in order to ensure that the complex amplitude of the electromagnetic field is modified from the input to the output scene simulating an unitary evolution of the state. Simulated unitary evolution allows to measure tensor products of local observables. Mean values of such quantities are experimentally evaluated from the intensity distribution of the field in the output image. We show that, depending on the encoded input state, correlated quantities calculated from the expected values of the observables violate a Clauser-Horne-Shimony-Holt Bell-type innequality.\\

In order to emphazise the role of interference effects in the Bell inequalities violation, we test our set-up in two different cases: In the first, the maximally entangled state $\left(\left|0\right\rangle_{A}\otimes\left|0\right\rangle_{B}+\left|1\right\rangle_{A}\otimes\left|1\right\rangle_{B}\right)/\sqrt{2}$ encoded in the input scene shows maximal violation meanwhile in the second, an incoherent superposition of $\left|0\right\rangle_{A}\otimes\left|0\right\rangle_{B}$ or $\left|1\right\rangle_{A}\otimes\left|1\right\rangle_{B}$ does not violate the Bell innequality mainly do to the absence of interference effects. Parametrical lost of coherence from maximally entangled to maximally mixed states has been simulated by means of an optical representation of a convex mixing of both type of states. In all cases, the experimental results of the simulations are in good agreement with the theoretical predictions based on quantum mechanics. This simulation clearly illustrates how the mechanism of Bell inequalities violation needs both resources: quantum entanglement and two-particle interference. Even though the subject of this paper constitutes a classical simulation of the quantum reality that only has an illustrative interpretation, it nicely demonstrates that classical optics is a useful tool for a deeper understanding of some fundamental aspects of quantum mechanics sometimes  considered to be hard and counterintuitive.
 
\section{Acknowledgments}

The authors are grateful to Prof. J. P. Paz and Prof. Marcos Sareceno for the stimulating and fruitful discussions and comments. This research was financed by the projects ANPCYT PICT 25373 and UBACYT X118.

\end{document}